\documentclass[aip,jcp,preprint,amsmath,amssymb]{revtex4-2}
\usepackage[version=4]{mhchem}
\usepackage{soulutf8}
\usepackage{graphicx}
\usepackage{xcolor}
\usepackage{bm}

\begin{document}

\title{Reaction-Path Statistical Mechanics of Enzymatic Kinetics}
\author{Hyuntae Lim}
\author{YounJoon Jung}
\affiliation{Department of Chemistry, Seoul National University, Seoul 08826, Korea}
\email{yjjung@snu.ac.kr}
\date{\today}

\begin{abstract}
We introduce a reaction-path statistical mechanics formalism based on the principle of large deviations to quantify the kinetics of single-molecule enzymatic reaction processes under the Michaelis-Menten mechanism, which exemplifies an out-of-equilibrium process in the living system. Our theoretical approach begins with the principle of \textit{equal a priori} probabilities and defines the reaction path entropy to construct a new nonequilibrium ensemble as a collection of possible chemical reaction paths. As a result, we evaluate a variety of path-based partition functions and free energies using the formalism of statistical mechanics. They allow us to calculate the timescales of a given enzymatic reaction, even in the absence of an explicit boundary condition that is necessary for the equilibrium ensemble. We also consider the large deviation theory under a closed-boundary condition of the fixed observation time to quantify the enzyme-substrate unbinding rates. The result demonstrates the presence of a phase-separation-like, bimodal behavior in unbinding events at a finite timescale, and the behavior vanishes as its rate function converges to a single phase in the long-time limit.
\end{abstract}

\maketitle

\section{Introduction}
The fundamental mathematical basis of statistical mechanics constitutes a probabilistic approach for describing complex systems,
starting from the principle of \emph{equal a priori probabilities} in the microcanonical ensemble\cite{chandler_introduction_1987,tuckerman_statistical_2010}.
In addition, various thermodynamic ensembles can be evaluated based on the mathematical formalism as an application of the \emph{large deviation principle}\cite{touchette_large_2009}.
This idea provides a concrete theoretical framework that is beneficial for considering numerous systems under equilibrium or near-equilibrium conditions,
whereas the thermodynamic ensemble approach is potentially invalid in far-from-equilibrium systems in which Boltzmann's postulate of ergodicity does not hold\cite{gallavotti_ergodicity_1995,evans_typicality_2016}.
However, the recent advances in nonequilibrium statistical mechanics introduced a framework\cite{gaspard_time-reversed_2004,lecomte_thermodynamic_2007,touchette_large_2009}, which poses a significant potential for enabling a comprehensive understanding of out-of-equilibrium systems via the same mathematical formalism as the thermodynamic ensemble approach\cite{garrahan_dynamical_2007,lecomte_thermodynamic_2007,hedges_dynamic_2009,speck_first-order_2012,weber_emergence_2013,budini_fluctuating_2014,vaikuntanathan_dynamic_2014,murugan_biological_2016,klymko_rare_2018,whitelam_large_2018}.
The construction of the dynamic ensemble as a collection of nonequilibrium \emph{trajectories} (or \emph{paths}) enables the calculation of out-of-equilibrium properties such as the number of dynamic \emph{events} in a given system, for instance, glass-forming liquids\cite{hedges_dynamic_2009,speck_first-order_2012}, spin-facilitated systems\cite{garrahan_dynamical_2007,budini_fluctuating_2014,jack_dynamical_2020,jack_ergodicity_2020}, kinetic networks\cite{vaikuntanathan_dynamic_2014,murugan_biological_2016}, and protein-folding pathways\cite{weber_emergence_2013}.

Based on these theoretical and numerical studies, we pursued the idea that the dynamic ensemble theory can be applied to realistic, out-of-equilibrium chemical reactions or biological processes in living systems. 
In this study, we propose a statistical ensemble approach to evaluate the timescales and unbinding events of Michaelis-Menten (MM) enzymatic kinetics, which is 
one of the most vital steps describing catalytic chemical reactions or biophysical processes\cite{johnson_original_2011},
\begin{equation}
    \ce{E + S <=>[$k_b$][$k_u$] C ->[$k_c$] E + P}.
    \label{eqn:mm_mechanism}
\end{equation}
Since its inception more than a century ago, MM kinetics have been studied, applied, and extended in numerous directions, both experimentally and theoretically \cite{cornish-bowden_one_2015,schnitzer_force_2000,qian_single-molecule_2002,asbury_kinesin_2003,oijen_single-molecule_2003,english_ever-fluctuating_2006,lomholt_manipulating_2007,grima_noise-induced_2009,golestanian_synthetic_2010,kim_macroscopic_2010,saha_nonrenewal_2011,li_aggregated_2013,aquino_chemical_2017,banerjee_accuracy_2017,park_nonclassical_2017,piephoff_conformational_2017,lin_electrochemistry_2018,barel_generality_2017,singh_statistical_2017,barel_integrated_2019,holehouse_stochastic_2020}.
Most recently, MM kinetics garnered significant attention from the perspective of single-molecule enzyme kinetics\cite{qian_single-molecule_2002,oijen_single-molecule_2003,kou_single-molecule_2005,english_ever-fluctuating_2006,cao_generic_2008,jung_novel_2010,yang_quantitative_2011,reuveni_role_2014,rotbart_michaelis-menten_2015,pal_landau-like_2019},
which has been previously analyzed from various theoretical frameworks,
such as the solution of simple linear differential equations\cite{qian_single-molecule_2002,kou_single-molecule_2005,english_ever-fluctuating_2006}, reaction time distribution method for more complex\cite{cao_generic_2008}, and non-Poissonian time distributions\cite{jung_novel_2010,yang_quantitative_2011,reuveni_role_2014,rotbart_michaelis-menten_2015,pal_landau-like_2019}.

The principal concept behind the MM mechanism is that there are two major reaction stages in the enzymatic catalysis process, as expressed in Eq.~(\ref{eqn:mm_mechanism}) 
\cite{johnson_original_2011}:
(i) reversible binding and unbinding reactions between the substrate (\ce{S}) and the enzyme (\ce{E}) molecules, \ce{E + S <=> C}, and
(ii) the irreversible catalytic reaction from the bound enzyme-substrate complex (\ce{C}) to the product (\ce{P}), \ce{C -> E + P}.
This concept guided us to envision the MM mechanism as a renewal process\cite{evans_stochastic_2020,gupta_work_2020}, and the unbinding reaction occurring in the stage (i) is essential for the entire process, as the unbinding reaction reverts a given system to its initial state, \ce{E + S}.

Thus, in the present study, we focus on quantifying the dynamic process of unbinding events, which is a measure of competition between the unbinding and catalysis reactions, based on the nonequilibrium thermodynamic ensemble and the large deviation theory\cite{lecomte_thermodynamic_2007,touchette_large_2009}.
Note that recent studies have suggested the importance of unbinding reactions and turnover behavior in several enzymatic processes\cite{reuveni_role_2014,rotbart_michaelis-menten_2015,pal_landau-like_2019}.
This proposed framework is  termed as the {\it reaction-path statistical mechanics} approach.
We initiate our formulation by defining a nonequilibrium ``microcanonical'' ensemble of equally probable reaction paths and its associated entropy for MM kinetics.
Thereafter, we introduce and evaluate various auxiliary nonequilibrium ensembles and their thermodynamic potentials, both directly and indirectly via the Legendre transform.
Furthermore, we proceed to explore the statistics of unbinding events and several timescales in a similar way as equilibrium thermodynamic calculations.
Lastly, we examine a phase-separation-like, bimodal behavior in the conditional probability density of unbinding events at a fixed observation time, and show that such a behavior vanishes in the long-time limit as the linear large deviation rate function converges to a single minimum.

\section{Theory}

\subsection{Reaction-Path Statistical Mechanics}

\subsubsection{Basic Scheme}

\begin{figure}
    \includegraphics[width = 8cm]{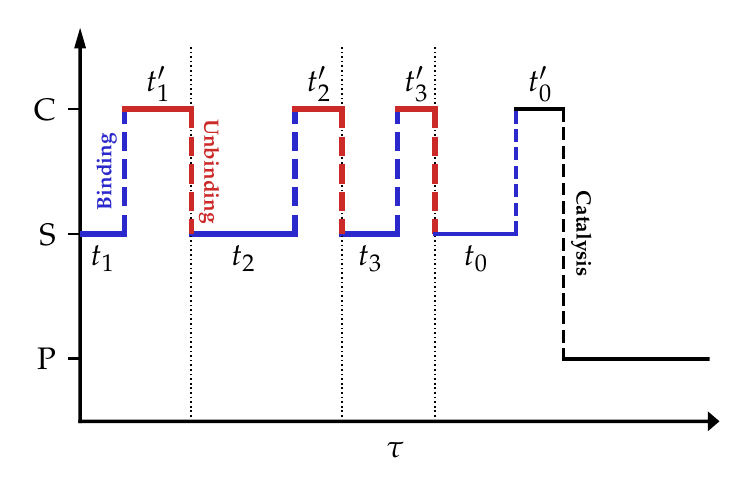}
    \caption{
        Schematic of Michaelis-Menten enzymatic reaction. Blue lines ($t_{i}$) denote enzyme-substrate binding (\ce{E + S -> C}), whereas red lines ($t_{i}^{'}$) correspond to unbinding reaction (\ce{C -> E + S}).
    }
    \label{fig:illust}
\end{figure}

The time-evolving stochastic dynamics of the given reaction system can be described by the chemical master equation (CME).
In this study, we consider a specific example of the MM kinetics in which the reaction initiates with only one enzyme (\ce{E}) and substrate (\ce{S}) molecule.
The equation introduces a Markov jump process for finite amounts of the reaction components with regard to a continuous time evolution.
Therefore, the time evolution of three components' probability, $p_\mathrm{S} (t)$, $p_\mathrm{C} (t)$, and $p_\mathrm{P} (t)$ could be expressed as the following master equation\cite{qian_single-molecule_2002,qian_chemical_2010}:
\begin{subequations}
    \begin{align}
        \dot{p}_\mathrm{S} (t) &= w_{u} p_\mathrm{C} (t) - w_{b} p_\mathrm{S} (t),
        \\
        \dot{p}_\mathrm{C}(t) &= w_{b} p_\mathrm{S} (t) - (w_{u} + w_{c}) p_\mathrm{C} (t),
        \\
        \dot{p}_\mathrm{P} (t) &= w_{c} p_\mathrm{C} (t),
    \end{align}
    \label{eqn:mm_single}
\end{subequations}
with the initial conditions of $p_\mathrm{S}(0) = 1$, $p_\mathrm{C}(0) = 0$, and $p_\mathrm{P}(0) = 0$.
In Eq.~(\ref{eqn:mm_single}), $w_{b} = k_{b} / V_{u}^2$, $w_{u} = k_{u} / V_{u}$, and $w_{c} = k_{c} / V_{u}$ denote the rate constants per unit volume $V_{u}$ of the stated system, and subscripts $b$, $u$, and $c$ denote the \textit{binding},  \textit{unbinding}, and  \textit{catalysis} stages, respectively.
As is illustrated in Fig. \ref{fig:illust}, We consider a single, completed reaction pathway $\mathrm{E} + \mathrm{S} \rightarrow \cdots \rightarrow \mathrm{E} + \mathrm{P}$ of Eq.~(\ref{eqn:mm_single}), involving $K$ unbinding events, $\mathrm{C} \rightarrow  \mathrm{E} + \mathrm{S}$.
Assuming that each kinetic step is a Markov process in nature and well described by the master equation,
we can associate each path with its own normalized probability $\rho [\{\mathrm{path}\}]$, which can be expressed as the successive products of Poissonian waiting-time distributions\cite{gaspard_time-reversed_2004,lecomte_thermodynamic_2007}:
\begin{equation}
    \begin{aligned}
        \rho [\{\mathrm{path}\}]
        = w_b e^{-w_b t_0} &\left( \prod_{i = 1}^{K} w_u e^{-(w_u + w_c) t'_{i}} w_b e^{-w_b t_{i}} \right)
        \\
        &\times w_c e^{-(w_u + w_c) t'_0},
    \end{aligned}
	\label{eqn:prob_base}
\end{equation}
where the parenthetical term denotes the survival probability of $\mathrm{C}$ after the occurrence of $K$ pairs of unbinding and binding events, and
the first and last terms correspond to the survival probabilities for the first binding and the final catalytic events, respectively.
Moreover, the time intervals $t_i$ and $t'_i$ denote the individual lifetimes of $\mathrm{S}$ and $\mathrm{C}$ at each reaction stage, respectively.
If the \emph{total lifetime} of each component is defined as the sum of individual lifetimes, $\tau_\mathrm{S} \equiv \sum_{i = 0}^{K} t_{i}$ and $\tau_\mathrm{C} \equiv \sum_{i = 0}^{K} t'_{i}$,
then Eq.~(\ref{eqn:prob_base}) can be simplified as
\begin{equation}
	\rho [\{\mathrm{path}\}]
	= w_\mathrm{b} w_\mathrm{c} (w_\mathrm{u} w_\mathrm{b})^{K} e^{-w_\mathrm{b} \tau_\mathrm{S}} e^{-(w_\mathrm{u} + w_\mathrm{c}) \tau_\mathrm{C}},
	\label{eqn:prob_simple}
\end{equation}
which only depends on the three path-dependent observables:
the number of unbinding events $K$, and the total lifetimes of the substrate molecule $\tau_\mathrm{S}$ and of the enzyme-substrate complex $\tau_\mathrm{C}$.
Upon considering these quantities $\{ \tau_\mathrm{S}, \tau_\mathrm{C}, K\}$ 
as a set of natural variables that characterize the reaction-path ensemble, we can collect all possible reaction paths bearing an equal probability.
Therefore, the concept of the \textit{nonequilibrium microcanonical} ensemble of reaction paths can be derived from the principle of equal a priori probabilities, and the corresponding entropy, $\mathcal{S}
\equiv  \ln \Omega (\tau_\mathrm{S}, \tau_\mathrm{C}, K)$, can be defined based on the equilibrium thermodynamic ensemble approach\cite{touchette_large_2009,tuckerman_statistical_2010}.

We evaluate the number of equal microscopic reaction paths, corresponding to \emph{micro-states} in equilibrium, with an identical set of $\{ \tau_\mathrm{S}, \tau_\mathrm{C}, K\}$ to calculate the entropy.
As the number of all the possible sets 
$\{ x_{0}, x_{1}, \cdots, x_{N} \}$ satisfying 
$x_{0} + x_{1} + \cdots + x_{N} = R$ and $0\leq x_{i} \leq R$
is equal to the volume of the geometric object defined by 
the hyperplane $x_{1} + \cdots + x_{N} = R$ with $0\leq x_{i}\leq R$, 
we can derive the exact expression of $\Omega$ as the product of two $K$-dimensional objects defined as $t_{1} + \cdots + t_{K} = \tau_\mathrm{S}$ and $t_{1}' + \cdots + t_{K}' = \tau_\mathrm{C}$:
\begin{subequations}
    \begin{align}
        {\Omega} (\tau_\mathrm{S}, \tau_\mathrm{C}, K) &= \frac{\tau_\mathrm{S}^{K}}{K!}\frac{\tau_\mathrm{C}^{K}}{K!},
        \label{eqn:microstates}\\
        {\cal S}(\tau_\mathrm{S},\tau_\mathrm{C},K) & 
        = \ln{\Omega} (\tau_\mathrm{S}, \tau_\mathrm{C}, K) 
        \simeq K\ln
        \left(e^2\tilde{\tau}_\mathrm{S}\tilde{\tau}_\mathrm{C}\right),
        \label{eqn:S_micro}
    \end{align}
    \label{eqn:microcanonical}
\end{subequations}
where $\tilde{\tau}_{\alpha}=\tau_{\alpha}/K (\alpha=\mathrm{E}$ or $\mathrm{C})$ 
denotes the average lifetime of the enzyme or of the complex, which is an intensive quantity.
The formulation of the path-entropy $\mathcal{S} (\tau_\mathrm{S}, \tau_\mathrm{C}, K)$ in Eq. (\ref{eqn:S_micro}) was simplified using Stirling's approximation.
Note that the nonequilibrium microcanonical partition function $\Omega (\tau_\mathrm{S}, \tau_\mathrm{C}, K)$ can be defined by the distinct number of possible arrangements for a mixture of two ideal gases, each consisting of an equal number of particles,
if we make the following identification $\tau\leftrightarrow V$ and $K\leftrightarrow N$, where $V$ and $N$ denote the volume and the number of particles, respectively.
Note that the path-entropy $\cal{S}$ is extensive in terms of the number of dynamic events $K$. 

The microcanonical ensemble introduced in Eq.~(\ref{eqn:microcanonical}) allows us to extend to alternative types of path ensembles based on the weighted sum (or integral) of the microcanonical partition function, $\Omega$:
\begin{subequations}
    \begin{align}
        \mathcal{Q}_{\mathrm{S}} 
        (\mu, \tau_\mathrm{C}, K)
        &=\int_0^\infty d {\tau_\mathrm{S}}
        e^{-\mu\tau_\mathrm{S}} \Omega(\tau_\mathrm{S},\tau_\mathrm{C},K) 
        = \frac{1}{\mu^{K + 1}} \frac{\tau_\mathrm{C}^{K}}{K!},
        \\
        \mathcal{Q}_{\mathrm{C}} 
        (\tau_\mathrm{S}, \nu, K)
        &=\int_0^\infty d {\tau_\mathrm{C}}
        e^{-\nu\tau_\mathrm{C}} \Omega(\tau_\mathrm{S},\tau_\mathrm{C},K) 
        = \frac{1}{\nu^{K + 1}} \frac{\tau_\mathrm{S}^{K}}{K!},
        \\         
        \mathcal{Q} (\mu, \nu, K)
        &=\int_0^\infty d {\tau_\mathrm{S}} e^{-\mu \tau_\mathrm{S}}
         \int_0^\infty d {\tau_\mathrm{C}} e^{-\nu \tau_\mathrm{C}} 
        \Omega(\tau_\mathrm{S},\tau_\mathrm{C},K)
        = \frac{1}{(\mu\nu)^{K + 1}},
        \\
        {\Xi} (\mu, \nu, \xi) 
        &= 
        \sum_{K=0}^{\infty} 
        \mathcal{Q} (\mu,\nu, K) e^{-\xi K}
        = \frac{1}{\mu \nu - e^{-\xi}},\label{eqn:all_intensive}
    \end{align}
    \label{eqn:partition}
\end{subequations}
where we introduced three intensive variables $\mu$, $\nu$, and $\xi$ as conjugate variables for their corresponding extensive quantities $\tau_\textrm{S}$, $\tau_\textrm{C}$, and $K$, respectively. Their physical meanings are discussed later in this paper. 
In addition, note that other ensembles can be defined further based on appropriate Laplace transforms.

Equation (\ref{eqn:partition}) evaluates the conditional probability densities of $\tau_\mathrm{S}$ and $\tau_\mathrm{C}$ in the $K$-fixed conditions and the marginal probability density for $K$ are:
\begin{subequations}
    \begin{align}
        \rho (\tau_\mathrm{S} | K) 
        &= 
        \frac{e^{-\mu\tau_\mathrm{S}}\Omega}{{\cal Q}_{\mathrm S}}
        =\frac{\mu^{K + 1} \tau_\mathrm{S}^{K}}{K!} e^{-\mu \tau_\mathrm{S}},
        \\
        \rho (\tau_\mathrm{C} | K) 
        &=
        \frac{e^{-\nu\tau_\mathrm{C}}\Omega}{{\cal Q}_{\mathrm C}}
        =\frac{\nu^{K + 1} \tau_\mathrm{C}^{K}}{K!} e^{-\nu \tau_\mathrm{C}},
        \\
        \rho (K)
        &=
        \frac{e^{-\xi K} \mathcal{Q}}{\Xi}
        =\left( 1 - \frac{e^{-\xi}}{\mu \nu} \right) \left( \frac{e^{-\xi}}{\mu \nu} \right)^{K}.
        \label{eqn:prob_k_fixed:marginal}
    \end{align}
    \label{eqn:prob_k_fixed}
\end{subequations}
Note that the two lifetimes $\tau_\mathrm{S}$ and $\tau_\mathrm{C}$ are mutually independent of each other.
As the enzymatic turnover time, $\tau_\mathrm{t}$, is the summation of $\tau_\mathrm{S}$ and $\tau_\mathrm{C}$,
its conditional probability density $\rho (\tau_\mathrm{t} | K)$ acquires a convoluted form of $\rho (\tau_\mathrm{S} | K)$ and $\rho (\tau_\mathrm{C} | K)$.
Although this convoluted form is too complex to resolve, it can be easily reduced in the Laplace domain.
Equation \ref{eqn:prob_k_fixed} allows us to evaluate the remaining probability densities such as $\rho(\tau_{S})$ or $\rho(\tau_{C})$ with Bayes' theorem, but their expectation values can be calculated from the partition functions in Eq. (\ref{eqn:all_intensive}) as nonequilibrium ensemble averages in a more straightforward manner:
\begin{subequations}
    \begin{align}
        \left< \tau_\mathrm{S} \right> = \frac{\nu}{\mu \nu - e^{-\xi}},
        \\
        \left< \tau_\mathrm{C} \right> = \frac{\mu}{\mu \nu - e^{-\xi}},
        \\
        \left< \tau_\mathrm{t} \right> \equiv \left< \tau_\mathrm{S} \right> + \left< \tau_\mathrm{C} \right> = \frac{\mu + \nu}{\mu \nu - e^{-\xi}},
        \\
        \left< K \right> = \frac{e^{-\xi}}{\mu \nu - e^{-\xi}}.
    \end{align}
    \label{eqn:ensemble_average}
\end{subequations}

Equation (\ref{eqn:ensemble_average}) demonstrates that the statistical mechanics approach reproduced the results from the solution of LDE\cite{qian_single-molecule_2002,kou_single-molecule_2005,english_ever-fluctuating_2006} for the conjugate variables $\mu = w_{b}$, $\nu = w_{u} + w_{c}$, and $\xi = -\ln w_{b} w_{u}$.
Another important thing we need to remark is that the last ensemble in Eq. (\ref{eqn:partition}), $\mu\nu \xi$ ensemble has no extensive variables.
In equilibrium statistical mechanics, an ensemble has at least one extensive variable because the variable bounds the given system\cite{allen_computer_1993}; if we would consider the $\mu PT$ ensemble, we would not evaluate $\left< N \right>$, $\left< V \right>$, or $\left< E \right>$ because the system has no boundary.
However, the evaluation of the mean values in Eq. (\ref{eqn:ensemble_average}) does not require any boundary conditions in experimental or theoretical approaches, so we can define a nonequilibrium ensemble involving only the intensive variables.

\subsubsection{Multi-Step Process}
In this section, we extend the proposed formalism for more complex reaction mechanisms.
Here, we consider a two-step enzymatic process that has an additional activation mechanism, $\ce{C <=> A}$, from the enzyme-substrate complex $\ce{C}$ to the activated complex, $\ce{A}$:
\begin{equation}
    \ce{
        E + S <=>[\it{b}][\it{u}] C <=>[\it{a}][\it{d}] A ->[\it{c}] P
    },
\end{equation}
where $a$ and $d$ denote reaction constants for \emph{activation} and \emph{deactivation}, while $b$, $u$, and $c$ are equivalent with $w_\mathrm{b}$, $w_\mathrm{u}$, and $w_\mathrm{c}$ in the previous section, respectively.
This two-step kinetics is also used to describe various reaction mechanisms, such as the gating dynamics of the enzyme molecule or the effect of an inhibition mechanism\cite{szabo_stochastically_1982,gopich_single-molecule_2007,qian_chemical_2010}.
If the given reaction system initiates at the same state, $\ce{E + S}$, the probability density of an individual reaction path towards to the turnover is given as the following form:
\begin{equation}
    \begin{aligned}
        \rho[\{ \text{path} \}] = 
        &b \exp \left( -b t^{(b)}_0 \right)
        \\
        &\times \left\{
            \prod_{i = 1}^{K_\text{u}}
            u \exp \left( -[a + u] t^{(u)}_{i} \right)
            b \exp \left( -b t^{(b)}_{i} \right)
        \right\}
        \\
        &\times \left\{
            \prod_{j = 1}^{K_\text{d}}
            a \exp \left( -[a + u] t^{(a)}_{j} \right)
            d \exp \left( -[c + d] t^{(d)}_{j} \right)
        \right\}
        \\
        &\times
        a \exp \left( -[a + u] t^{(a)}_{0} \right)
        c \exp \left( -[c + d] t^{(c)}_{0} \right)
        \\
        =&
        bac (ub)^{K_{u}} (da)^{K_{d}}
        e^{-b \tau_{\text{S}}}
        e^{-(a + u) \tau_{\text{C}}}
        e^{-(c + d) \tau_{\text{A}}},
    \end{aligned}
\end{equation}
that leads us to define two new extensive variables: the lifetime of activated state $\ce{A}$, $\tau_{{A}}$ and the number of deactivation transitions $\ce{A -> C}$, $K_\text{d}$.
We also change the notation of unbinding count from $K$ to $K_\text{u}$ for clarity.
Then the three of total lifetimes $\tau_{{S}}$, $\tau_{{C}}$, and $\tau_{{S}}$ can be expressed as:
\begin{subequations}
    \begin{align}
        \tau_\text{S} &=
        t^{(b)}_{0} + \left( t^{(b)}_{1} \cdots + t^{(b)}_{K_\text{u}} \right),
        \\
        \tau_\text{C} &=
        t^{(a)}_{0} +
        \left( t^{(u)}_{1} + \cdots + t^{(u)}_{K_\text{u}}
        + t^{(a)}_{1} + \cdots + t^{(a)}_{K_\text{d}} \right),
        \\
        \tau_\text{A} &=
        t^{(c)}_{0} + \left( t^{(d)}_{1} + \cdots + t^{(d)}_{K_\text{d}} \right).
    \end{align}
\end{subequations}
Now we can consider that the microscopic number of equivalent turnover trajectories, $\Omega_{2}$ of the two-step kinetics as a product of three volumes under the three hyperplanes whose dimensions are $K_\text{u}$, $K_\text{u} + K_\text{d}$, and $K_\text{d}$, respectively. $\Omega_{2}$ also requires an additional binomial mixing term that accounts for two competitive reactions, $\ce{C <=> E + S}$ and $\ce{C <=> A}$:
\begin{equation}
    \begin{aligned}
        \Omega_{2}(\tau_{S}, \tau_{C}, \tau_{A}, K_\text{u}, K_\text{d})
        &=
        \frac{\tau_{S}^{K_\text{u}}}{K_\text{u}!}
        \frac{\tau_{C}^{K_\text{u} + K_\text{d}}}{(K_\text{u} + K_\text{d})!}
        \frac{\tau_{A}^{K_\text{d}}}{K_\text{d}!}
        \cdot
        \frac{(K_\text{u} + K_\text{d})!}{K_\text{u}! K_\text{d}!}
        \\
        &=
        \Omega(\tau_{S}, \tau_{C}, K_\text{u})
        \Omega(\tau_{C}, \tau_{A}, K_\text{d}).
    \end{aligned}
\end{equation}
The mixing term makes $\Omega_{2}$ a simple product of two different microcanonical partition functions of single-step processes that share a timescale, $\tau_\mathrm{C}$.
Thus, we can proceed to evaluate the corresponding ensemble partition functions, $\mathcal{Q}_{2}$ and $\Xi_{2}$:
\begin{subequations}
    \begin{align}
        \mathcal{Q}_{2} (\mu, \nu, \gamma, K_\text{u}, K_\text{d}) &= \frac{1}{(\mu \nu)^{K_\text{u} + 1}} \frac{1}{(\nu \gamma)^{K_\text{d} + 1}} \frac{(K_\text{u} + K_\text{d})!}{K_\text{u}! K_\text{d}!},
        \\
        \Xi_{2} (\mu, \nu, \gamma, \xi, \zeta) &= \frac{1}{\mu\nu\gamma - (\gamma e^{-\xi} + \mu e^{-\zeta})},
    \end{align}
    \label{eqn:partition_twostep}
\end{subequations}
where $(\gamma, \zeta)$ and $(\tau_\text{A}, K_\text{d})$ are conjugate to each other, while $(\mu, \nu, \xi)$ are taken from the single step kinetics, respectively.

For the general model of an arbitrary $N$-step sequential process given as:
\begin{equation}
    \ce{
        C^{(0)} <=>[\it{b}_{0}][\it{u}_{0}] C^{(1)} <=>[\it{b}_{1}][\it{u}_{1}]
    }
    \cdots
    \ce{
        <=>[\it{b}_{\it{N}-1}][\it{u}_{\it{N}-1}] C^{(\it{N})} ->[\it{b}_{\it{N}}] P,
    }
\end{equation}
the microcanonical ensemble of the given mechanism, $\Omega_{N}$ can be evaluated as the successive products of single-step reactions, while the corresponding partition function $\mathcal{Q}_{N}$ has two product terms that consist of the single-step $\mathcal{Q}$ and the binomial mixing term:
\begin{subequations}
    \begin{align}
        \Omega_{N} (\tau_{0}, \cdots, \tau_{N}, K_{1}, \cdots, K_{N}) &=
        \prod_{n = 1}^{N} \Omega (\tau_{n - 1}, \tau_{n}, K_{n}),
        \\
        \mathcal{Q}_{N} (\mu_{0}, \cdots, \mu_{N}, K_{1}, \cdots, K_{N}) &=
        \prod_{n = 1}^{N} \mathcal{Q} (\mu_{i - 1}, \mu_{i}, K_{i})
        \prod_{n = 1}^{N - 1} \frac{(N_{n} + N_{n + 1})!}{N_{n}! N_{n + 1}!},
    \end{align}
\end{subequations}
where $\mu_{n}$ is the conjugate variable which biases $\tau_{n}$.
It is challenging to evaluate the general form of the intensive partition function $\Xi_{N}$ exactly, however, it can shown that the following recurrence relation holds (see {\bfseries Appendix A}):
\begin{equation}
    \begin{aligned}
        &\Xi_{N+1}(\{\mu_{n}\}, \{K_{n}\})
        =
        &\frac{\Xi_{N-1}(\{\mu_{n}\}, \{K_{n}\}) \Xi_{N}(\{\mu_{n}\}, \{K_{n}\})}{\mu_{N + 1} \Xi_{N-1}(\{\mu_{n}\}, \{K_{n}\}) - e^{-\xi_{N + 1}}\Xi_{N}(\{\mu_{n}\}, \{K_{n}\})},
    \end{aligned}
    \label{eqn:partition_recurrence}
\end{equation}
where $\xi_{n}$ biases $K_{n}$.
Using Eq. (\ref{eqn:partition_recurrence}), we can calculate $\Xi_{N}$ since we have explicit expressions of $\Xi_{N}$ for $N = 1$ (Eq. (\ref{eqn:partition})) and $N = 2$ (Eq. (\ref{eqn:partition_twostep})).

\begin{figure}
    \includegraphics[width = 16cm]{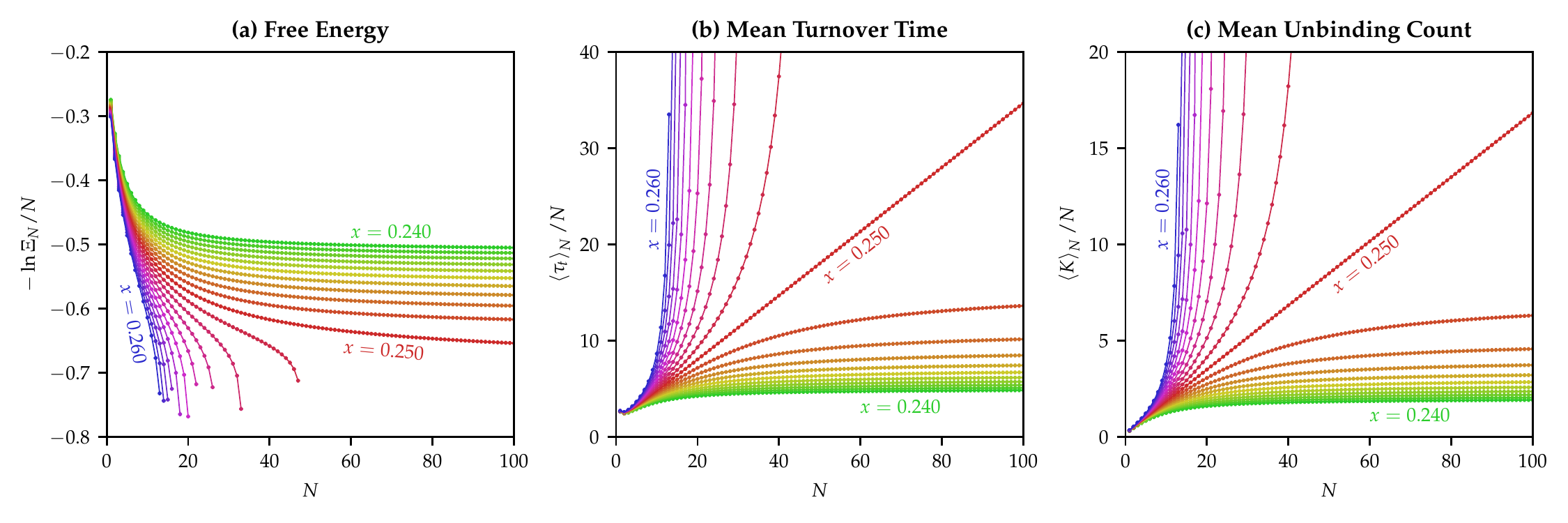}
    \caption{Step size dependences of (a) free energy, (b) mean turnover time, and (c) mean unbinding counts for the homogeneous $N$-step model, where $\mu$ was set to $1$ and $x$ denotes $e^{-\xi} / \mu^{2}$.}
    \label{fig:homogeneous_model}
\end{figure}

If we consider a specific example where the reaction system's ensemble consists of homogeneous intensive variables, $\mu_{n} = \mu$ and $\xi_{n} = \xi$ for all $n$'s, then the exact expression of $\Xi_{N}$ is given as:
\begin{equation}
    \Xi_{N}(\mu, \xi) =
    \frac{1}{\mu^{N + 1}}
    \frac{2^{N + 2} y}{(1 + y)^{N + 2} - (1 - y)^{N + 2}}
    \label{eqn:partition_homogeneous}
\end{equation}
where $x \equiv e^{-\xi}/\mu^{2}$ and $y \equiv \sqrt{1 - 4x}$, respectively.
The second fractional term Eq. (\ref{eqn:partition_homogeneous}) can be expanded as a polynomial form of $x$, for examples:
\begin{equation}
    \begin{gathered}
        \Xi_{3} (\mu, \xi) = \mu^{-4}(1 - 3x + x^{2})^{-1},
        \\
        \Xi_{4} (\mu, \xi) = \mu^{-5}(1 - 4x + 3x^{2})^{-1},
        \\
        \Xi_{5} (\mu, \xi) = \mu^{-6}(1 - 5x + 6x^{2} - x^{3})^{-1},
        \\
        \Xi_{6} (\mu, \xi) = \mu^{-7}(1 - 6x + 10x^{2} - 4x^{3})^{-1},
        \\
        \cdots,
    \end{gathered}
\end{equation}
and the coefficient of $x^{n}$ term in $\Xi_{N} (\mu, \xi)$ can be generalized as $(N - n + 1)!/n!(N - 2n + 1)!$ with alternating signs. In Fig. (\ref{fig:homogeneous_model}) we depicted $N$-dependence behavior of the free energy $-\ln \Xi_{N}$, and two ensemble averages $\left< \tau_\textrm{t} \right>_{N} \equiv -\partial_{\mu} \ln \Xi_{N}$ and $\left< K \right>_{N} \equiv -\partial_{\xi} \ln \Xi_{N}$ for the homogeneous $N$-step model, evaluated from Eq. (\ref{eqn:partition_recurrence}).
For $0 < x < 1/4$ (and $0 < y < 1$), we can easily find an asymptotic limit $- N^{-1} \ln \Xi_{N} \sim \ln [\mu (1 + y) / 2]$ which results in $\mathcal{O} (N)$ dependencies of $\left< \tau_{t} \right>_{N}$ and of $\left< K \right>_{N}$ for $N \gg 1$.
The system's behavior with $N$ demonstrates a drastic change at $x = 1/4$ (and $y = 0$):
both $\left< \tau_{t} \right>_{N}$ and $\left< K \right>_{N}$ shares $\mathcal{O}(N^{2})$ dependence for large $N$, as we depicted as the red lines in Fig. \ref{fig:homogeneous_model}.
The recurrence formula in Eq. (\ref{eqn:partition_recurrence}) became unstable where the value of $x$ is greater than $1/4$, yielded an unphysical, diverging partition function $\Xi_{N}$ for a certain $N$, as well as an absence of the corresponding asymptotic limit.
That is to say, the homogeneous kinetics is not expandable for an arbitrary size of $N$ for $e^{-\xi} / \mu^{2} > 1/4$.

\subsection{Ensemble of Fixed Observation Time}

In Sec. II A, we have shown that it is feasible to quantify the enzymatic kinetics in a language of statistical mechanics.
The evaluation was done with opened (or unbound) conditions for the timescales and unbinding events, which allowed us to calculate the turnover timescales from the ensemble of intensive variables.
However, in a certain theoretical or experimental scenario, sampling the reaction paths with arbitrary \textit{observation time} $\tau_\mathrm{obs}$ might be more appropriate, instead of using the fixed number of enzyme-substrate unbinding events $K$\cite{garrahan_dynamical_2007,budini_fluctuating_2014} due to various reasons, for example, experimental limitations or side reactions.
Under the given condition of fixed observation time, we consider \emph{completed} reaction paths as well as sampled \emph{incomplete} reaction paths remaining in the $\left| \mathrm{S} \right>$ or $\left| \mathrm{C} \right>$ at the given observation time.
Thus, the time evolution of the system can be described using the propagator $\mathbb{U} (\tau_\mathrm{obs})$,
\begin{equation}
    \left| p (\tau_\mathrm{obs}) \right>
        = \mathbb{U} (\tau_\mathrm{obs}) \left| p (0) \right>
        = \exp (\tau_\mathrm{obs} \mathbb{W}) \left| p (0) \right>,
\end{equation}
%
where $\left| p(t) \right> = p_\mathrm{S} (t) \left| \mathrm{S} \right> + p_\mathrm{C} (t) \left| \mathrm{C} \right> + p_\mathrm{P} (t) \left| \mathrm{P} \right>$ involves all possible stochastic paths in the time interval $(0, \tau_\mathrm{obs})$.
Therefore, we can decompose the propagator into the operators of conditional probabilities for unbinding events $K$, $\mathbb{U}(\tau_\mathrm{obs}) = \sum_{K} \mathbb{P} (K | \tau_\mathrm{obs})$, and the corresponding conditional probability is $P(K | \tau_\mathrm{obs}) = \left< \mathrm{e} \right| \mathbb{P} (K | \tau_\mathrm{obs}) \left| \mathrm{S} \right>$ where $\left| \mathrm{e} \right> = \left| \mathrm{S} \right> + \left| \mathrm{C} \right> + \left| \mathrm{P} \right>$ denotes the \textit{projection state}.

Recent studies have reported that the probability density of \emph{dynamic events} may show a multi-modal behavior, which signifies more than two dynamical phases in systems exhibiting heterogeneous or glassy dynamics\cite{garrahan_dynamical_2007,budini_fluctuating_2014,jung_novel_2010,hedges_dynamic_2009,speck_first-order_2012,weber_emergence_2013,vaikuntanathan_dynamic_2014,murugan_biological_2016,whitelam_large_2018,klymko_rare_2018}.
Similarly, the Michaelis-Menten mechanism displays heterogeneous kinetics in its unbinding events and results in the inactive (unbinding-poor) and active (unbinding-rich) trajectories, in terms of $P(K | \tau_\mathrm{obs})$ at a timescale near the mean turnover time $\left< \tau_\textrm{t} \right>$.
In the remaining part of this paper, we describe how MM kinetics can exhibit such a behavior using two distinct theoretical approaches.

The first approach does not directly yield $P(K | \tau_\mathrm{obs})$, but it uses the formalism of the large deviation principle to evaluate the moment-generating function of $K$ with the corresponding conjugate variable $s$\cite{touchette_large_2009,garrahan_dynamical_2007,budini_fluctuating_2014}:
\begin{equation}
	Z(s, \tau_\mathrm{obs})
    = \sum_{K = 0}^{\infty} e^{-sK} P(K | \tau_\mathrm{obs})
    \sim e^{\tau_\mathrm{obs} \phi(s)},
	\label{eqn:sensemble_partition}
\end{equation}
where the $n$-th derivative of $\ln Z(s, \tau_\mathrm{obs})$ provides the $n$-th moment of the unbinding events $K$ at a fixed observation time.
In addition, the $n$-th cumulant can be evaluated from the $n$-th derivative of the cumulant generating function, $\phi (s, \tau_\mathrm{obs}) = \tau_\mathrm{obs}^{-1} \ln Z (s, \tau_\mathrm{obs})$.
Although the exact analytical expression of $P(K | \tau_\mathrm{obs})$ may not exist, we could calculate the partition function from the matrix product states\cite{garrahan_dynamical_2007}:
\begin{equation}
	Z(s, \tau_\mathrm{obs}) = \left< \mathrm{e} \right| \exp(\tau_\mathrm{obs} e^{-s} \mathbb{W}_\mathrm{m} + \tau_\mathrm{obs} \mathbb{W}_\mathrm{r}) \left| \mathrm{S} \right>.
	\label{eqn:partition_matrix}
\end{equation}
The first matrix term in Eq. (\ref{eqn:partition_matrix}), $\mathbb{W}_\mathrm{m}$ corresponds to the \emph{monitored} transition that is the interest of the partition function.
When we count the amount of unbinding reaction $\mathrm{C} \rightarrow \mathrm{S}$, $\mathbb{W}_\mathrm{m}$ will be equal to $w_{u} \left| \mathrm{S} \right> \left< \mathrm{C} \right|$.
The second term $\mathbb{W}_\mathrm{r}$ denotes the rest of the reactions in the master operator $\mathbb{W} - \mathbb{W}_\mathrm{m}$.
In the ``thermodynamic'' limit, where the $\tau_\mathrm{obs}$ is sufficiently long,
the largest eigenvalue of the matrix $\mathbb{W}_{s} = e^{-s} \mathbb{W}_\mathrm{m} + \mathbb{W}_\mathrm{r}$ yields the large deviation function $\phi (s) = \lim_{\tau_\mathrm{obs} \rightarrow \infty} \phi (s, \tau_\mathrm{obs})$:
\begin{equation}
	\phi (s) = 
	\begin{cases}
	0
	& s > s_\mathrm{c},
	\\
	\frac{-(\mu + \nu) + \sqrt{(\mu - \nu)^2 + 4 e^{-s -\xi}}}{2}
	& s \leq s_\mathrm{c}.
	\label{eqn:ld_limit},
	\end{cases}
\end{equation}
where
\begin{equation}
    s_\mathrm{c} = -\xi - \ln \mu \nu = - \ln (1 + w_c / w_u).
\end{equation}
The detailed mathematical evaluations of $\phi(s, \tau_\mathrm{obs})$ and $\phi(s)$ are available at \textbf{Appendix C}.
The second part of Eq. (\ref{eqn:ld_limit}) is smaller than zero for an $s$ greater than $s_\mathrm{c}$, which indicates that $\phi (s)$ displays singularity at $s = s_\mathrm{c}$.

As described in Eq. (\ref{eqn:ld_limit}), $s_\mathrm{c}$ always yields a negative value for any proportions of $w_\mathrm{c}$ over $w_\mathrm{u}$, and the mean rate of unbinding at the point $s_{c}$ is $\left< K / \tau_\mathrm{obs} \right>_{s = s_\mathrm{c}} = w_b (1+ e^{s_c} \left< K \right>_{s = 0})^{-1}$, where $\left< K \right>_{s = 0}$ denotes the result of Eq. (\ref{eqn:ensemble_average}).
A singularity in $\phi (s)$ may imply the coexistence of multiple dynamical phases, but it is not the case for the proposed model since it exhibits a linear large deviation rate function\cite{oono_large_1989,touchette_large_2009,corberi_probability_2019}:
\begin{equation}
    I(k) \sim -\lim_{\tau_\mathrm{obs \rightarrow \infty}} \tau_\mathrm{obs}^{-1}\ln P(K)
    = -k \ln p,
    \label{eqn:rate}
\end{equation}
where $P(K)$ is the result of Eq. (\ref{eqn:prob_k_fixed:marginal}),
$p \equiv w_\mathrm{u}/(w_\mathrm{u} + w_\mathrm{c})$, and $k \equiv K/\tau_\mathrm{obs}$.
Thus, $I(k)$ exhibits only one minimum at $k = 0$, and the system generates a finite $\left< K \right>$ in the limit of $\tau_\mathrm{obs} \rightarrow \infty$.

For a non-zero value of $s$, we can find Doob's $h$-transform of $\mathbb{W}_{s}$ obtained with the first left-eigenvector of $\mathbb{W}_{s}$ (See \textbf{Appendix C})\cite{chetrite_variational_2015,ray_exact_2018}:
\begin{equation}
    \tilde{\mathbb{W}}_{s} =
    \begin{pmatrix}
        -w_\mathrm{b}  &w_\mathrm{u}e^{-s}      &0  \\
        w_\mathrm{b}   &-w_\mathrm{u}-w_\mathrm{c}   &0  \\
        0   &w_\mathrm{c}+(1-e^{-s})w_\mathrm{u}      &0
    \end{pmatrix}.
    \label{eqn:doob}
\end{equation}
The tilted master operator in Eq. (\ref{eqn:doob}) also results in a finite $\left< K\right>_{s}$ where $s>s_\mathrm{c}$, that makes $\left<k\right>_{s}=0$ in the long-time limit.
The \emph{tilted rate} of $\ce{C}\ce{->}\ce{P}$, $w_\mathrm{c}+(1-e^{-s})w_\mathrm{u}$ becomes zero at $s=s_\mathrm{c}$,
which means the system is unable to proceed to the turnover and exhibit a non-zero $\left< k \right>_{s=s_\mathrm{c}}$.
In contrast, there are no trajectories which exhibit $\left< k \right> = 0$ unless the reaction initiates at the absorbing state $\left| \mathrm{P} \right>$\cite{oono_large_1989}, but this is not the case of what we have supposed.
Consequently, the singularity of $\phi(s)$ does not suggest a coexistence of different dynamical phases in the long-time limit\cite{whitelam_large_2018,whitelam_varied_2021}.

\begin{figure}
    \includegraphics[width = 16cm]{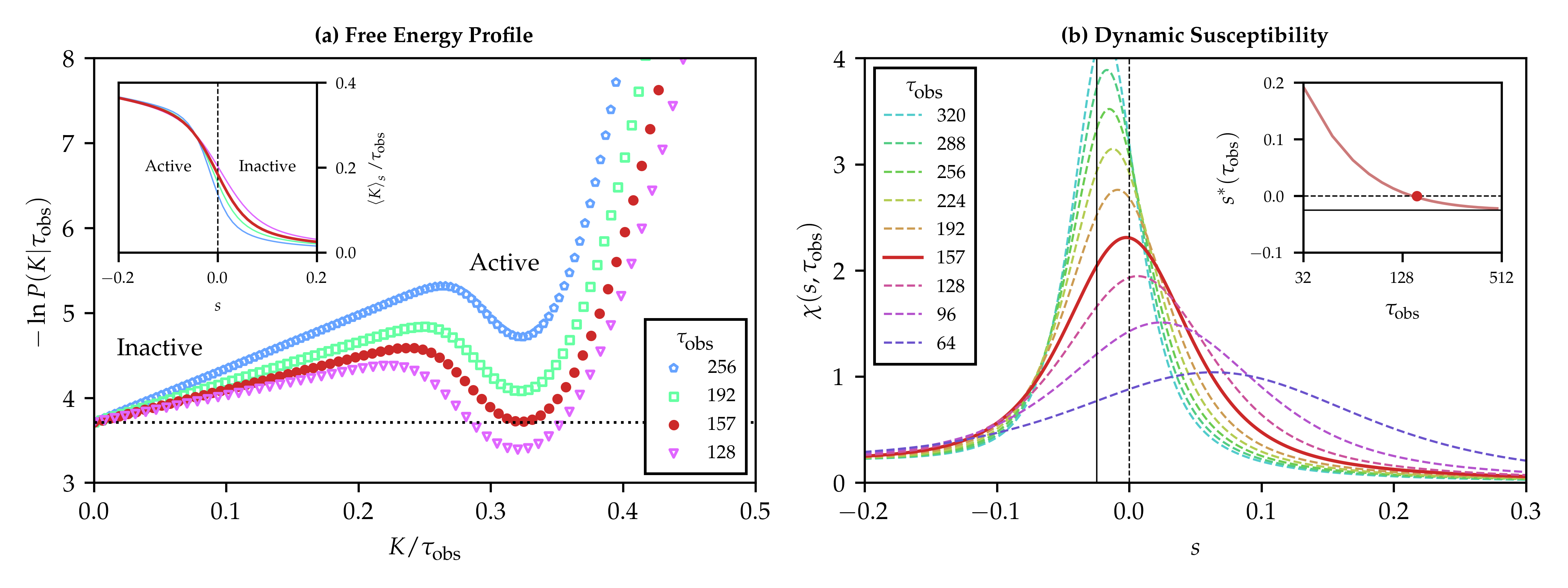}
    \caption{
        (a) Landau free energy profiles, $-\ln P(K | \tau_\mathrm{obs})$ for unbinding events $K$ at given observation times $\tau_\mathrm{obs}$ and
        (b) dynamic susceptibility curves $\chi_{k} (s, \tau_\mathrm{obs})$ that were evaluated from Eq. (\ref{eqn:partition_matrix}).
        Both the red plots in (a) and the red line in (b) indicates phase-coexistence-like behavior at the same observation time scale.
        Rate constants: $w_b = 0.5$, $w_u = 1.0$, and $w_c = 0.025$.
    }
    \label{fig:susceptibility}
\end{figure}

However, at the finite timescale near the turnover time $\left< \tau_\mathrm{t} \right>$, the partition function $Z(s, \tau_\mathrm{obs})$ and its corresponding probability density $P(K| \tau_\mathrm{obs})$
exhibits much more complicated behavior, as shown in Eq. (\ref{eqn:ld_general_form}).
The variation of the dynamic susceptibility, $\chi_{k} (s, \tau)$, which is the second derivative of $\phi (s, \tau_\mathrm{obs})$, with the observation time is presented in Fig. \ref{fig:susceptibility}, which further denotes the amount of fluctuations of unbindings for a given observation time, $k \equiv K / \tau_\mathrm{obs}$.
The dynamic susceptibility reaches its maximum value at the point $s = s^{*}$, which separates the reaction paths for a finite observation time into two distinct sets: the active ($s < s^{*}$) and inactive ($s > s^{*}$) trajectories.
Note that the conjugate variable $s$ is virtual, and its real physical meaning is not well known.
The only established knowledge is that $s$ has to be regarded as zero when one samples the reaction paths of the system in ordinary, unbiased conditions.
Therefore, the crossover timescale $\tau^{*}$, where $s^{*} (\tau)$ becomes zero, should be determined as we depicted in Fig. \ref{fig:susceptibility}(b).

\begin{figure}
    \includegraphics[width = 16cm]{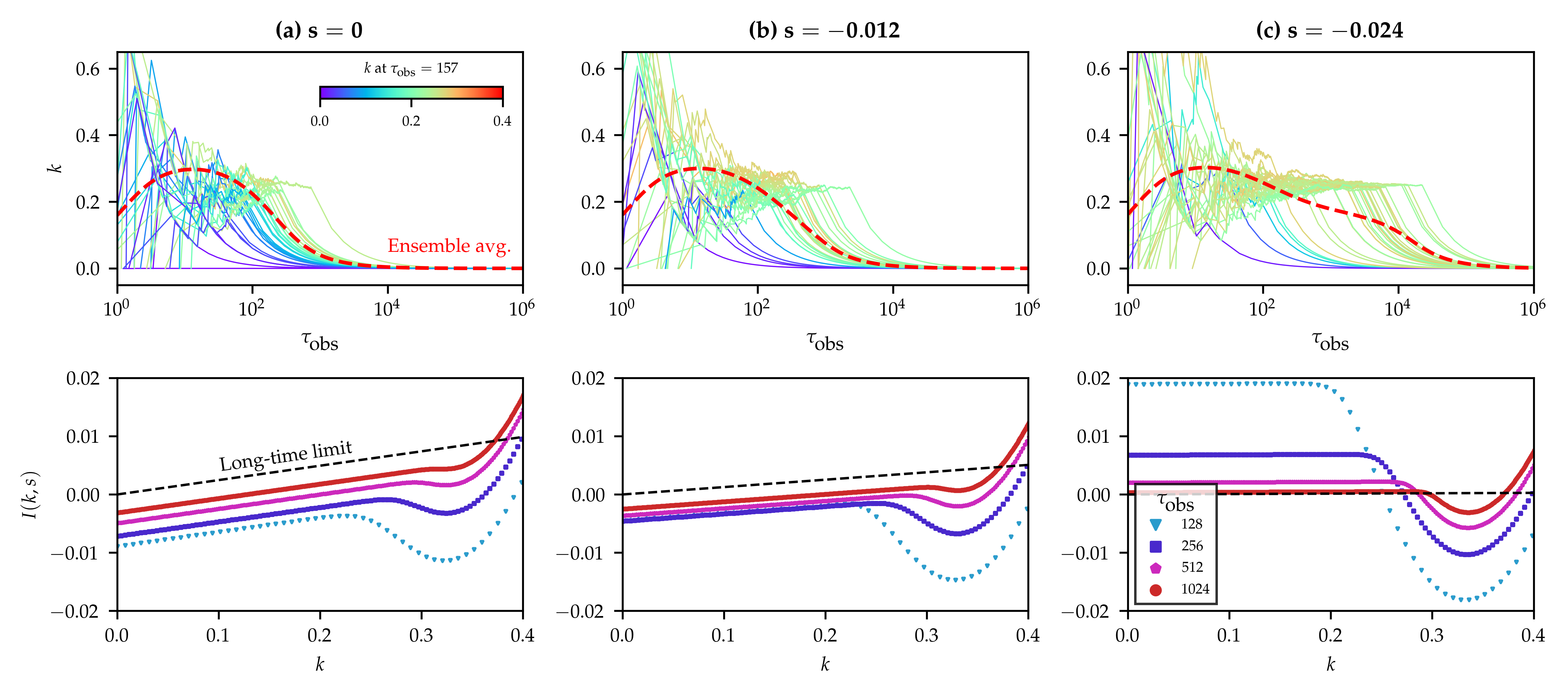}
    \caption{
        Upper figures depict the kinetic Monte Carlo trajectories for (a) unconstrained model at $s = 0$, (b) $s=-0.012$, and (c) $s = -0.024$ where $s_{c} \simeq -0.02469$, respectively. The solid lines denote individual trajectories where the color corresponds to $k = K / \tau_\textrm{obs}$ at $\tau_\textrm{obs} = 157$,
        while the red dashed line is their ensemble averaged value, respectively. The three figures at bottom depicts the large deviation rate function $I(s, k)$ at finite timescales and the LD limit.
    }
    \label{fig:tilted}
\end{figure}

In this study, we argue that the behavior of dynamic susceptibility indicates a dynamical phase-coexistence-like behavior at a certain, finite timescale $\tau^{*}$ near the mean turnover time.
Simultaneously, we have to confirm that $P (K | \tau_\mathrm{obs})$ exhibits the same trend at a given $\tau^{*}$.
Thus, we propose an alternative, numerically exact method to evaluate $P (K | \tau_\mathrm{obs})$ for a more comprehensive comparison, which does not require any limit conditions such as $\tau_\mathrm{obs} \gg 1$ or $K \gg 1$, in contrast to Eq. (\ref{eqn:ld_limit}).
The key idea of the method is the decomposition of the propagator $\mathbb{U} (\tau_\mathrm{obs})$ into the $K$-th order terms of $\mathbb{W}_\mathrm{m}$, as the term accounts for the number of unbinding reactions.
In general, the propagator has a matrix exponential form of the master operator. Therefore, the decomposed term can be expressed in the following form:
\begin{equation}
	\mathbb{P} (K | \tau_\mathrm{obs}) = \sum_{n = 0}^{\infty} \frac{\tau_\mathrm{obs}^{K + n}}{(K + n)!} \mathbb{O}(K, n),
	\label{eqn:prob_conditional_matrix}
\end{equation}
where $\mathbb{O} (K, n)$ is the $K$-th order term of $\mathbb{W}_\mathrm{m}$ from the polynomial $(\mathbb{W}_\mathrm{m} + \mathbb{W}_\mathrm{r})^{K + n}$ and can be calculated from the recurrence formula, $\mathbb{O} (K, n) = \mathbb{W}_\mathrm{m} \mathbb{O} (K - 1, n) + \mathbb{W}_\mathrm{r} \mathbb{O} (K, n - 1)$ (See {\bfseries Appendix B}).
Fig. \ref{fig:susceptibility}-(a) illustrates the Landau-like free energy profile of $- \ln P (K | \tau_\mathrm{obs})$, numerically calculated using Eq. (\ref{eqn:prob_conditional_matrix}) with an arbitrarily chosen cutoff value $n_\mathrm{max}=4096$ in the infinite sum.
As shown in Fig. \ref{fig:susceptibility}(a), $- \ln P (K | \tau_\mathrm{obs})$ has its global minimum at $K / \tau_\mathrm{obs} = 0$ for $\tau_\mathrm{obs} > 157$ while $K / \tau_\mathrm{obs} \simeq 0.32$ for $\tau_\mathrm{obs} \leq 157$.
This is a consistent result with the other graphs in Fig. \ref{fig:susceptibility}, including the inset of Fig. \ref{fig:susceptibility}(a) and both the main figure and inset of (b), which use the exact analytical form of the moment-generating function $Z(s, \tau_\mathrm{obs})$, evaluated at Eq. (\ref{eqn:ld_general_form}) in \textbf{Appendix C}.
As $\tau_\mathrm{obs}$ increases, turnover trajectories are gradually accumulated and the bimodal behavior in $P(K|\tau_\mathrm{obs})$ fades away; consequently, $-\tau_\mathrm{obs}^{-1} \ln P (K | \tau_\mathrm{obs})$ converges to the large deviation rate function $I(k)$, which exhibits a single minimum at $k = 0$, as we discussed in Eq. (\ref{eqn:rate}).
For a non-zero $s$ values, we depicted the convergence  behavior of $-\tau_\mathrm{obs}^{-1} \ln P_{s} (K | \tau_\mathrm{obs})$ to $I(k, s)$ in Fig. \ref{fig:tilted}, obtained from Eq. (\ref{eqn:prob_conditional_matrix}) and the tilted operator Eq. (\ref{eqn:doob}).
The figure also depicts the individual trajectories of the tilted dynamics that were sampled with Kinetic Monte Carlo (KMC) simulations.

\begin{figure}
    \centering
    \includegraphics[width = 8cm]{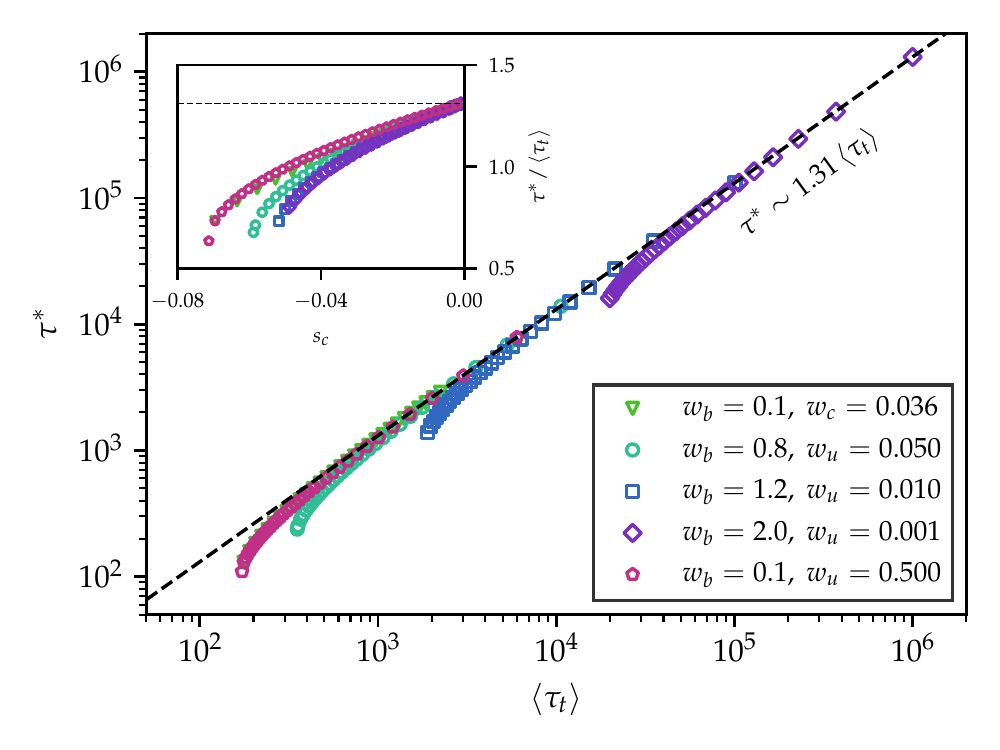}
	\caption{Relation between mean-turnover times, $\left< t_\mathrm{t} \right>$ and active-inactive transition times,$\tau^{*}$.
	Two of three reaction constants were fixed, whereas remaining one is varied.
    Black dashed line clarifies that all datasets represent linearly correlated tendency, approximately $d \tau^{*} / d \left< t_\mathrm{t} \right> \simeq 1.32$ in large turnover time scale.}
    \label{fig:timescale}
\end{figure}

Finally, the mean values of the turnover times $\tau_\mathrm{t}$ versus numerically calculated $\tau^{*}$ at various binding, unbinding, and catalysis rates, as depicted in Fig. \ref{fig:timescale}.
Consequently, strong linear correlations were determined between $\tau^{*}$ and $\left< \tau_\mathrm{t} \right>$.
In addition, each data set represents the case in which two of the three rate constants are fixed, and the remaining one is varied;
the linear relation, $d \tau^{*} / d \left< \tau_\mathrm{t} \right> \sim 1.3$, becomes apparent for $w_u \gg w_c$.
As the catalysis stage is irreversible and a single reaction is over, the number of unbinding events of the given path ceases to increase.
This results in the population of the inactive paths continually increasing with the observation time, and \textit{vice versa} for the active paths.
In the thermodynamic (or long-time) limit where $\tau_\mathrm{obs} \rightarrow \infty$, only the inactive paths survive and $P (K | \tau_\mathrm{obs})$ converges to $\rho (K)$ for a sufficiently long time period, which we discussed earlier.
Therefore, the value of $s_\mathrm{c}$ in the long-time limit can be calculated from the convergence of Eq. (\ref{eqn:sensemble_partition}), $\sum_{K=0}^{\infty} e^{-sK} \rho (K)$.
This preference for the inactive trajectories in a long observation time scale of the system would cause a phase-coexistence-like behavior at $\tau^{*}$, in case that the set of reaction paths shows an active $P (K | \tau_\mathrm{obs})$ at a short time scale.
Understandably, the logic can vary depending on the relative rate constants;
it is unable to determine a positive $\tau^{*}$ if the rate of catalysis, $w_{c}$, is sufficiently greater than the rate of unbinding, $w_{u}$.
In this scenario, $s^{*} (\tau_\mathrm{obs})$ yields a negative value even at a very short observation time $\tau_\mathrm{obs}$ at all instances,
and the system keeps exhibiting inactive trajectories from the beginning to the end of the reactions since the process does not generate a sufficient amount of $K$ over all possible trajectories.
This principle provides a lower boundary as depicted in Fig. \ref{fig:timescale}.

\section{Conclusions}
In the current study, we demonstrate that mathematical formalism for statistical mechanics is suitable for quantifying single-molecule enzymatic kinetics under the Poissonian MM mechanism.
In addition, three variables, namely, the number of unbinding events, and total lifetimes of the substrate and of the enzyme-substrate complex, were used in combination with the principle of \textit{equal a priori} probabilities under nonequilibrium conditions, and the reaction path entropy was defined.
Based on this concept, we successfully evaluated three statistical ensembles of the out-of-equilibrium process: microcanonical ($\tau_\mathrm{S}\tau_\mathrm{C}K$), canonical ($\mu\tau_\mathrm{C}K$) and ($\tau_\mathrm{S}\nu K$), and grand canonical ($\mu\nu K$) ensembles.
Moreover, the conjugate intensive variables in these ensembles, $\mu$ and $\nu$ bias the statistical weights of trajectories, with the lifetimes of components $\tau_\textrm{S}$ and $\tau_\textrm{C}$, respectively.
The definition of a single reaction path was formulated, where $\nu$ and $\mu$ denote the escaping ratios of the Markov process.
The thermodynamic relations between the nonequilibrium ensembles yielded the probability distributions of several important reaction-time scales.
The results obtained from the reaction path statistical mechanics reproduced previous results based on the mean-field theory.
We also extended the proposed approach to a more complex, generalized multi-step enzymatic kinetics.
The homogeneous model showed $\mathcal{O}(N)$ dependences for both $\left< \tau_\mathrm{t} \right>_{N}$ and $\left< K \right>_{N}$ where $e^{-\xi} / \mu^{2} < 1/4$, whereas they exhibited $\mathcal{O}(N^{2})$ at $e^{-\xi} / \mu^{2} = 1/4$, which is the stable limit of the model.

Furthermore, various theoretical or experimental scenarios were considered, as we extended our results to a fixed observation time, $\tau_\mathrm{obs}$.
We evaluated the Bayesian statistics and performed numerical calculations to demonstrate the phase-separation-like behavior of the enzymatic reaction. In particular, the number of unbinding events per observation time, $k = K / \tau$, is used as an order parameter.
These two trajectory collections were named as inactive (unbinding-poor) and active (unbinding-rich) trajectories.
While the system becomes inactive for a sufficiently long observation time,
a crossover from the active to the inactive behaviors may appear during the reaction process, 
depending on the combination of the reaction rate constants.
In addition, the timescale $\tau^{*}$, where such a crossover appears, apparently exhibits an approximately linear relationship with the average value of enzymatic turnover time, $\left< \tau_\mathrm{t} \right>$.

As the unbinding of enzyme-substrate is evidently essential in the kinetics of complex enzymatic processes,
the current work proposes a potential way to quantify the dynamical behaviors of systems under the MM mechanism.
We will extend our study to general models, especially the non-Poisson (or heterogeneous) enzymatic reaction process of the enzymatic reaction process.
In addition, we expect that our study on the nonequilibrium ensemble theory can be applied to various systems in out-of-equilibrium conditions.

\section*{Acknowledgments}
This work was supported by Samsung Science and Technology Foundation (grant number SSTF-BA1601-11).
We are grateful to an anonymous referee who pointed out Refs. \citenum{oono_large_1989} and \citenum{whitelam_varied_2021}, and suggested the tilted dynamics calculation.

\bibliography{references}

\appendix

\section{N-step Partition function}

We have to evaluate the following summation to obtain $N$-step partition function, $\Xi_{N}$:
\begin{equation}
    q_{N}
    \equiv \sum_{K_{N}} \cdots \sum_{K_{1}}
    x_{1}^{K_{1}} \cdots x_{N}^{K_{N}}
    \frac{(K_{1} + K_{2})!}{K_{1}! K_{2}!}
    \cdots
    \frac{(K_{N-1} + K_{N})!}{K_{N-1}! K_{N}!}
    \label{eq:nstep_partition_sum}
\end{equation}
where $x_{i} \equiv e^{-\xi_{i}} / \mu_{i - 1} \mu_{i}$ and $\Xi_{N} = q_{N} / \mu_{0} \cdots \mu_{N}$.
If we use the binomial series,
\begin{equation}
    \sum_{N = 0}^{\infty} \frac{(M + N)!}{M! N!} p^{N}
    = \left( \frac{1}{1 - p} \right)^{M + 1},
\end{equation}
then the sum in Eq. (\ref{eq:nstep_partition_sum}) can be proceeded step-by-step:
\begin{equation}
    \begin{aligned}
        q_{N}
        &=
        \sum_{K_{N}} \cdots \sum_{K_{2}}
        \frac{1}{1 - x_{1}}
        \left(
            \frac{x_{2}}{1 - x_{1}}
        \right)^{K_{2}}
        \frac{(K_{2} + K_{3})!}{K_{2}! K_{3}!}
        \cdots
        \frac{(K_{N-1} + K_{N})!}{K_{N-1}! K_{N}!}
        \\
        &=
        \sum_{K_{N}} \cdots \sum_{K_{3}}
        \frac{1}{1 - x_{1}}
        \frac{1}{1 - \frac{x_{2}}{1 - x_{1}}}
        \left(
            \frac{x_{3}}{1 - \frac{x_{2}}{1 - x_{1}}}
        \right)^{K_{3}}
        \frac{(K_{3} + K_{4})!}{K_{3}! K_{4}!}
        \cdots
        \frac{(K_{N-1} + K_{N})!}{K_{N-1}! K_{N}!}
        \\
        &=
        \sum_{K_{N}} \cdots \sum_{K_{3}}
        \frac{1}{1 - x_{1} - x_{2}}
        \left(
            \frac{(1 - x_{1}) x_{3}}{1 - x_{1} - x_{2}}
        \right)^{K_{3}}
        \frac{(K_{3} + K_{4})!}{K_{3}! K_{4}!}
        \cdots
        \frac{(K_{N-1} + K_{N})!}{K_{N-1}! K_{N}!}
        \\
        &\cdots
        \\
        &=
        \sum_{K_{N}} \cdots \sum_{K_{n}}
        q_{n - 1}
        \left(
            \frac{q_{n - 1}}{q_{n - 2}} x_{n}
        \right)^{K_{n}}
        \frac{(K_{n} + K_{n + 1})!}{K_{n}! K_{n + 1}!}
        \cdots
        \frac{(K_{N-1} + K_{N})!}{K_{N-1}! K_{N}!}
        \\
        &\cdots
        \\
        &=
        \sum_{K_{N}}
        q_{N - 1}
        \left( \frac{q_{N - 1}}{q_{N - 2}} x_{N} \right)^{K_{N}}
        =
        \frac{q_{N - 1} q_{N - 2}}{q_{N - 2} - q_{N - 1} x_{N}}
    \end{aligned}
\end{equation}
Finally we obtain the recurrence relation in Eq. (\ref{eqn:partition_recurrence})
\begin{equation}
    \Xi_{N + 1} = \frac{q_{N + 1}}{\mu_{0} \cdots \mu_{N + 1}}
    = \frac{\Xi_{N - 1} \Xi_{N}}{\mu_{N + 1} \Xi_{N - 1} - e^{-\xi_{N}} \Xi_{N}}
\end{equation}

\section{Propagator decomposition}

In the present appendix, we will let $b \equiv w_{b}$, $u \equiv w_{u}$, and $c \equiv w_{c}$ in consideration of legibility.
Then the master operator $\mathbb{W}$, which governs the time-evolving kinetics of given catalytic reaction system, is expressed as the following 3$\times$3 matrix\cite{qian_chemical_2010}:
\begin{equation}
    \mathbb{W} =
    \begin{pmatrix}
        -b  &u      &0  \\
        b   &-u-c   &0  \\
        0   &c      &0
    \end{pmatrix},
\end{equation}
where its bases are $\left| \mathrm{S} \right> \equiv \left| 1 \right>$, $\left| \mathrm{ES} \right> \equiv \left| 2 \right>$, and $\left| \mathrm{P} \right> \equiv \left| 3 \right>$, respectively.
Since we count the number of unbinding transitions $\left| \mathrm{ES} \right> \rightarrow \left| \mathrm{S} \right>$,
we decompose $\mathbb{W}$ into two separate operators: the transition that we will monitor, $\mathbb{W}_\mathrm{m}$ and the rest of them, $\mathbb{W}_\mathrm{r} \equiv \mathbb{W} - \mathbb{W}_\mathrm{m}$\cite{touchette_large_2009},
\begin{subequations}
    \begin{gather}
        \mathbb{W}_\mathrm{m} =
        \begin{pmatrix}
            0   &u  &0  \\
            0   &0  &0  \\
            0   &0  &0
        \end{pmatrix},
        \\
        \mathbb{W}_\mathrm{r} =
        \begin{pmatrix}
            -b  &0      &0  \\
            b   &-u-c   &0  \\
            0   &c      &0
        \end{pmatrix}.
    \end{gather}
\end{subequations}
Given the definition of propagator $\left| p (\tau_\mathrm{obs}) \right> = \mathbb{U} (\tau_\mathrm{obs}) \left| p (0) \right>$ that consists of all \emph{possible} stochastic trajectories with equal time lengths (or \emph{observation time}) $\tau_\mathrm{obs}$, we can regard the propagator as sum of all trajectories which involve $K$ \emph{monitored} transitions,
\begin{equation}
    \mathbb{U} (\tau_\mathrm{obs}) = \sum_{K = 0}^{\infty} \mathbb{P} (K | \tau_\mathrm{obs}).
\end{equation}
Here, the operator $\mathbb{P} (K | \tau_\mathrm{obs})$ should consist of $K$-th order terms of $\mathbb{W}_\mathrm{m}$ since it corresponds to the normalized conditional probability of $K$ at fixed $\tau_\mathrm{obs}$,
\begin{equation}
    P (K | \tau_\mathrm{obs}) = \left< \mathrm{e} \right| \mathbb{P} (K | \tau_\mathrm{obs}) \left| \mathrm{S} \right>,
\end{equation}
where $\left| \mathrm{e}\right> \equiv \left| \mathrm{S}\right> + \left| \mathrm{C}\right> + \left| \mathrm{P}\right>$ and $\left| p(0) \right> = \left| \mathrm{S} \right>$.
Because of the propagator $\mathbb{U} (\tau_\mathrm{obs})$ has an exponential form,
We can evaluate the series expansion and sort its elements according to the order of $\mathbb{W}_\mathrm{m}$:
\begin{equation}
    \begin{aligned}
        \mathbb{U} (\tau_\mathrm{obs})
            \equiv& \exp (\tau_\mathrm{obs} \mathbb{W})\\
        =&1 + \tau_\mathrm{obs}(\mathbb{W}_\mathrm{m} + \mathbb{W}_\mathrm{r}) + \frac{1}{2} \tau_\mathrm{obs}^{2}(\mathbb{W}_\mathrm{m} + \mathbb{W}_\mathrm{r})^{2} + \frac{1}{6} \tau_\mathrm{obs}^{3}(\mathbb{W}_\mathrm{m} + \mathbb{W}_\mathrm{r})^{3} + \cdots
        \\
        =&
        \left( 1 + \tau_\mathrm{obs} \mathbb{W}_\mathrm{r} + \frac{1}{2} \tau_\mathrm{obs}^{2} \mathbb{W}_\mathrm{r}^{2} + \frac{1}{6} \tau_\mathrm{obs}^{3} \mathbb{W}_\mathrm{r}^{3} + \cdots \right)
        \\
        &+
        \left(\tau_\mathrm{obs} \mathbb{W}_\mathrm{m}
        + \frac{1}{2} \tau_\mathrm{obs}^{2} \left( \mathbb{W}_\mathrm{m}\mathbb{W}_\mathrm{r} + \mathbb{W}_\mathrm{r}\mathbb{W}_\mathrm{m} \right)
        + \frac{1}{6} \tau_\mathrm{obs}^{3} \left( \mathbb{W}_\mathrm{m} \mathbb{W}_\mathrm{r}^{2} + \mathbb{W}_\mathrm{r} \mathbb{W}_\mathrm{m} \mathbb{W}_\mathrm{r} + \mathbb{W}_\mathrm{r}^{2} \mathbb{W}_\mathrm{m} \right) + \cdots \right)
        \\
        &+ \left(
        \frac{1}{2}\tau_\mathrm{obs}^{2} \mathbb{W}_\mathrm{m}^{2} + \frac{1}{6} \tau_\mathrm{obs}^{3} \left( \mathbb{W}_\mathrm{m}^{2} \mathbb{W}_\mathrm{r} + \mathbb{W}_\mathrm{m} \mathbb{W}_\mathrm{r} \mathbb{W}_\mathrm{m} + \mathbb{W}_\mathrm{r} \mathbb{W}_\mathrm{m}^{2} \right) + \cdots
        \right)
        \\
        &+ \cdots
        \\
        =&
        \sum_{K = 0}^{\infty} \sum_{n = 0}^{\infty} \frac{\tau_\mathrm{obs}^{K + n}}{(K + n)!} \mathbb{O} (\mathbb{W}_\mathrm{m}^{K}, \mathbb{W}_\mathrm{r}^{n}),
    \end{aligned}
\end{equation}
which gives the exact analytical expression of $\mathbb{P} (K | \tau_\mathrm{obs})$,
\begin{equation}
    \mathbb{P} (K | \tau_\mathrm{obs})
        = \sum_{n = 0}^{\infty} \frac{\tau_\mathrm{obs}^{K + n}}{(K + n)!} \mathbb{O} (\mathbb{W}_\mathrm{m}^{K}, \mathbb{W}_\mathrm{r}^{n}).
    \label{eqn:prob_series}
\end{equation}
Here, $\mathbb{O} (\mathbb{W}_\mathrm{m}^{K}, \mathbb{W}_\mathrm{r}^{n})$ is sum of $K$-th order terms of $\mathbb{W}_\mathrm{m}$ from the 
polynomial $\left( \mathbb{W}_\mathrm{m} + \mathbb{W}_\mathrm{r} \right)^{K + n}$
and it has a recurrence relation,
\begin{equation}
    \mathbb{O} (\mathbb{W}_\mathrm{m}^{K}, \mathbb{W}_\mathrm{r}^{n})
    = \mathbb{W}_\mathrm{m} \mathbb{O} (\mathbb{W}_\mathrm{m}^{K - 1}, \mathbb{W}_\mathrm{r}^{n})
    + \mathbb{W}_\mathrm{r} \mathbb{O} (\mathbb{W}_\mathrm{m}^{K}, \mathbb{W}_\mathrm{r}^{n - 1}),
    \label{eqn:recurrence}
\end{equation}
with two initial conditions $\mathbb{O} (\mathbb{W}_\mathrm{m}^{K}, \mathbb{W}_\mathrm{r}^{0}) = \mathbb{W}_\mathrm{m}^{K}$ and $\mathbb{O} (\mathbb{W}_\mathrm{m}^{0}, \mathbb{W}_\mathrm{r}^{n}) = \mathbb{W}_\mathrm{r}^{n}$, respectively.
Eqs. (\ref{eqn:prob_series}) and (\ref{eqn:recurrence}) enable us to calculate $P(K | \tau_\mathrm{obs})$ in a numerical way with a maximum cutoff value of $n$.

\section{Calculation of the partition function}

From the large deviation theory, the partition function (or \emph{moment-generating} function) of $K$ can be evaluated via canonical sum over conditional probabilities of $K$ with a biasing field $s$\cite{lecomte_thermodynamic_2007,garrahan_dynamical_2007,touchette_large_2009}:
\begin{equation}
    \begin{aligned}
        Z(s, \tau_\mathrm{obs})
            &= \sum_{K = 0}^{\infty} e^{-s K} P (K | \tau_\mathrm{obs})
            \\
            &= \left< \mathrm{e} \right| \sum_{K = 0}^{\infty} e^{-s K} \mathbb{P} (K | \tau_\mathrm{obs}) \left| \mathrm{S} \right>
    \end{aligned}
\end{equation}
Here, we adopt the series expression of $\mathbb{P} (K | \tau_\mathrm{obs})$ from Eq. (\ref{eqn:prob_series}) to evaluate the partition sum:
\begin{equation}
    \begin{aligned}
        \sum_{K = 0}^{\infty} e^{-s K} \mathbb{P} (K | \tau_\mathrm{obs})
            &= \sum_{K = 0}^{\infty} \sum_{n = 0}^{\infty}
                \frac{\tau_\mathrm{obs}^{K + n}}{(K + n)!} e^{-sK} \mathbb{O} ( \mathbb{W}^{K}_\mathrm{m}, \mathbb{W}^{n}_\mathrm{r})
            \\
            &= \sum_{K = 0}^{\infty} \sum_{n = 0}^{\infty}
            \frac{\tau_\mathrm{obs}^{K + n}}{(K + n)!} \mathbb{O} ( (e^{-s} \mathbb{W}_\mathrm{m})^{K}, \mathbb{W}^{n}_\mathrm{r})
            \\
            &= \exp(\tau_\mathrm{obs} \mathbb{W}_{s}),
    \end{aligned}
\end{equation}
where $\mathbb{W}_{s} \equiv e^{-s} \mathbb{W}_\mathrm{m} + \mathbb{W}_\mathrm{r}$.
If the operator $\mathbb{W}_{s}$ is diagonalizable with $\mathbb{W}_{s} = \mathbb{T} \mathbb{J} \mathbb{T}^{-1}$, we can calculate the partition function as
\begin{equation}
    Z(s, \tau_\mathrm{obs}) = \left< \mathrm{e}\right| \mathbb{T} \exp (\tau_\mathrm{obs} \mathbb{J}) \mathbb{T}^{-1} \left| \mathrm{S} \right>,
\end{equation}
and each element of the diagonal matrix $\mathbb{J}$ corresponds to an eigenvalue of $\mathbb{W}_{s}$.
That is to say, the partition function is a linear combination of three eigenvalues: $\lambda_{1} (s)$, $\lambda_{2} (s)$, and $\lambda_{3} (s)$, respectively.
\begin{equation}
    Z(s, \tau_\mathrm{obs}) = C_{1}(s) e^{\tau_\mathrm{obs} \lambda_{1}(s)} + C_{2}(s) e^{\tau_\mathrm{obs} \lambda_{2}(s)} + C_{3}(s) e^{\tau_\mathrm{obs} \lambda_{3}(s)}.
\end{equation}

\subsection{Large deviation limit}

In the large deviation limit where $\tau_\mathrm{obs}$ goes to infinite, only the largest eigenvalue of $\mathbb{W}_{s}$ will survive and be the large deviation function $\phi(s)$\cite{garrahan_dynamical_2007,budini_fluctuating_2014}:
\begin{equation}
    \phi(s)
        = \lim_{\tau_\mathrm{obs} \rightarrow \infty} \tau_\mathrm{obs}^{-1} \ln Z(s, \tau_\mathrm{obs})
        =\sup_{i} \lambda_{i} (s).
\end{equation}
We can calculate the eigenvalues from the characteristic polynomial of $\mathbb{W}_{s}$, which is a cubic equation:
\begin{equation}
    \begin{aligned}
        \det (\mathbb{W}_{s} - \lambda \mathbb{I}) &=
        \begin{vmatrix}
            -b-\lambda  &e^{-s} u           &0  \\
            b           &-(u + c) - \lambda &0  \\
            0           &c                  &-\lambda
        \end{vmatrix}
        \\
        &= -\lambda \left( \lambda^{2} + (b + u + c) \lambda + (bc + bu - e^{-s} bu) \right)
        \\
        &= -\lambda \left( \lambda^{2} + (\mu + \nu) \lambda + (\mu \nu - e^{-s} e^{\xi}) \right)
    \end{aligned}
\end{equation}
Here, we use three intensive variables of the nonequilibrium ensemble: $\mu \equiv b$, $\nu \equiv u + c$, and $\xi \equiv -\ln bu$, respectively.
Then three eigenvalues from $\det (\mathbb{W}_{s} - \lambda \mathbb{I}) = 0$ are expressed as:
\begin{subequations}
    \begin{align}
        \lambda_{1} &= 0,
        \\
        \lambda_{2} &= \frac{-(\mu + \nu) - \sqrt{(\mu - \nu)^{2} + 4 e^{-s-\xi}}}{2},
        \\
        \lambda_{3} &= \frac{-(\mu + \nu) + \sqrt{(\mu - \nu)^{2} + 4 e^{-s-\xi}}}{2},
    \end{align}
    \label{eqn:eigenvalues}
\end{subequations}
respectively. The term in the square root, $(\mu - \nu)^{2} + 4 e^{-s-\xi}$ is always positive so we have three real eigenvalues.
Here, $\lambda_{2}$ is not the largest one since it is always negative, and we have to find the boundary of $\lambda_{3} \geq \lambda_{1}$ to get the largest value:
\begin{equation}
    \begin{gathered}
        \sqrt{(\mu - \nu)^{2} + 4 e^{-s-\xi}} \geq \mu + \nu
        \\
        e^{-s - \xi} \geq \mu\nu,
        \\
        s \leq -\xi - \ln \mu\nu.
    \end{gathered}
\end{equation}
So the coexistence point is $s_{c} \equiv -\xi -\ln \mu\nu = -\ln(1 + c/u)$, and it always has negative value.
The large deviation function, $\phi (s)$ has a singularity at $s_{c}$:
\begin{equation}
    \phi(s) = 
    \begin{cases}
        \lambda_{1}(s) ~~~~~~ s > s_{c}
        \\
        \lambda_{3}(s) ~~~~~~ s \leq s_{c}.
    \end{cases}
\end{equation}
Thus, we have two mean values of unbinding density over observation time, $\lim_{\tau_\mathrm{obs} \rightarrow \infty}\left< k \right>_{s} = \left< K \right>_{s, \tau_\mathrm{obs}} / \tau_\mathrm{obs}$ at the point:
the inactive trajectories,
\begin{equation}
    \lim_{s \downarrow s_{c}} \left< k \right>_{s}
        = \left. - \frac{\partial \lambda_{1} (s)}{\partial s} \right|_{s \downarrow s_{c}}
        = 0,
    \label{eqn:k_inactive}
\end{equation}
and active trajectories,
\begin{equation}
    \begin{aligned}
        \lim_{s \uparrow s_{c}} \left< k \right>_{s}
            &= \left. - \frac{\partial \lambda_{3} (s)}{\partial s} \right|_{s \uparrow s_{c}}
            \\
            &= \left(\frac{1}{\mu} + \frac{1}{\nu}\right)^{-1}.
    \end{aligned}
    \label{eqn:k_active}
\end{equation}

\subsection{General form}

\begin{figure*}
    \includegraphics[width = 16cm]{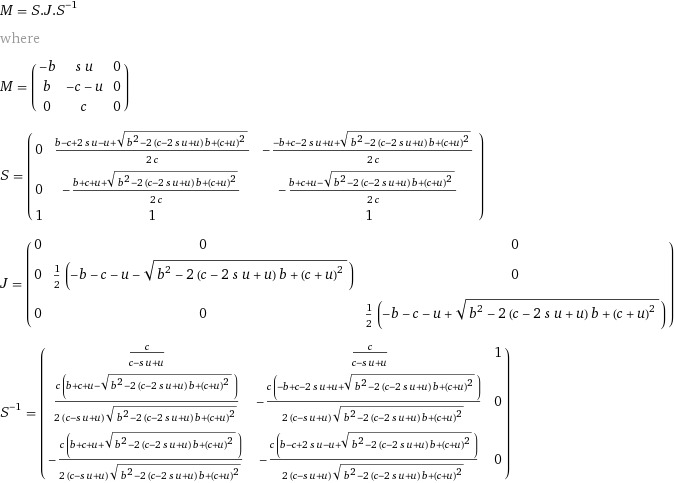}
    \caption{Diagonalization procedure for the modified master operator, $\mathbb{W}_{s}$. The symbol $s$ in this figure stands for $e^{-s}$.}
    \label{fig:wolfram}
\end{figure*}

\begin{figure*}
    \includegraphics[width = 16cm]{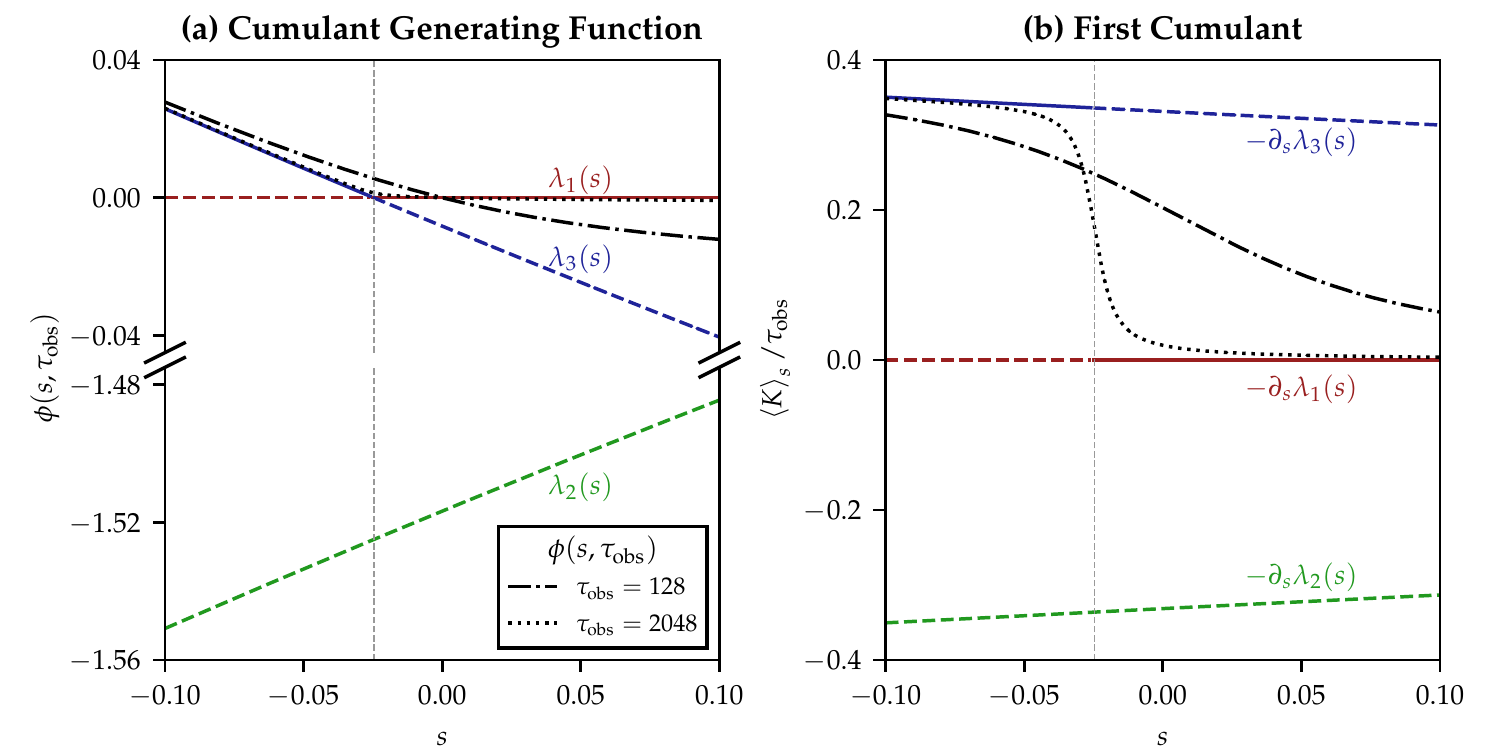}
    \caption{Analytical behaviours of three eigenvalues $\lambda_{i} (s)$ and their first derivatives, $-\partial_{s} \lambda_{i}(s)$, respectively. Black lines depict the partition function, $Z(s, \tau_\mathrm{obs})$ at $\tau_{obs} =$ 128 and 2048.}
    \label{fig:ld}
\end{figure*}

We consider more general cases for a finite observation time, $\tau_\mathrm{obs}$. That is to say, we have to evaluate
\begin{equation}
    Z(s, \tau_\mathrm{obs}) = \left< e \right| \exp (\tau_\mathrm{obs} \mathbb{W}_{s}) \left| \mathrm{S} \right>,
\end{equation}
where $\left| e \right> \equiv \left| \mathrm{S} \right> + \left| \mathrm{ES} \right> + \left| \mathrm{P} \right>$.
Because the matrix $\mathbb{W}_{s}$ is diagonalizable as shown in Fig. \ref{fig:wolfram}, we can evaluate the general form of the partition function with
\begin{equation}
    \begin{aligned}
        Z(s, \tau_\mathrm{obs}) &= \left< e \right| \exp (\tau_\mathrm{obs} \mathbb{W}_{s}) \left| \mathrm{S} \right>
        \\
        &= \left< e \right| \mathbb{S} \exp(\tau_\mathrm{obs} \mathbb{J}) \mathbb{S}^{-1} \left| \mathrm{S} \right>.
    \end{aligned}
\end{equation}
The exact analytical expression of $Z(s, \tau_\mathrm{obs})$ is:
\begin{equation}
    \begin{aligned}
        Z(s, \tau_\mathrm{obs})
        &= \frac{\mu \nu - e^{-\xi}}{\mu \nu - e^{-s -\xi}}
        e^{\tau_\mathrm{obs} \lambda_{1} (s)}
        \\
        &+ \frac{(e^{-s} - 1)e^{-\xi}}{\mu \nu - e^{-s -\xi}} \cdot \frac{(\mu + \nu) + \sqrt{(\mu - \nu) + 4 e^{-s - \xi}}}{\sqrt{(\mu - \nu) + 4 e^{-s - \xi}}}
        e^{\tau_\mathrm{obs} \lambda_{2} (s)}
        \\
        &+ \frac{(1 - e^{-s})e^{-\xi}}{\mu \nu - e^{-s -\xi}} \cdot \frac{(\mu + \nu) + \sqrt{(\mu - \nu) + 4 e^{-s - \xi}}}{\sqrt{(\mu - \nu) + 4 e^{-s - \xi}}}
        e^{\tau_\mathrm{obs} \lambda_{3} (s)},
        \label{eqn:ld_general_form}
    \end{aligned}
\end{equation}
where $\lambda_{1} (s)$, $\lambda_{2} (s)$, and $\lambda_{3} (s)$ are results of Eq. (\ref{eqn:eigenvalues}).
Fig. \ref{fig:ld} shows analytical behaviors of the three eigenvalues, $Z(s, \tau_\mathrm{obs})$, and their first derivatives, respectively.

\section{Tilted Dynamics}
The biased reaction trajectories with a non-zero value of $s$ can be sampled by accomplishing Doob's $h$-transform on $\mathbb{W}_{s}$, which can be expressed as\cite{ray_exact_2018,chetrite_variational_2015}:
\begin{equation}
    \tilde{\mathbb{W}}_{s} (x, x') = \psi_{s}(x) \mathbb{W}_{s} (x,x') \psi_{s}^{-1} (x') - \phi(s),
\end{equation}
where $\psi_{s}$ is the corresponding eigenvector of the scaled cumulant generation function, $\phi(s)$. 
$x$ and $x'$ denote the matrix (or vector) indices. Here, if we use the first left eigenvector of $\mathbb{W}_{s}$, that is the first row of $S^{-1}$ in Fig. \ref{fig:wolfram}:
\begin{equation}
    \psi_{s} =
    \left(
        \frac{c}{c + (1-e^{-s})u}, \frac{c}{c + (1-e^{-s})u}, 1
    \right),
\end{equation}
then we can achieve $\tilde{\mathbb{W}}_{s}$ with the following form:
\begin{equation}
    \tilde{\mathbb{W}}_{s} =
    \begin{pmatrix}
        -b  &ue^{-s}      &0  \\
        b   &-u-c   &0  \\
        0   &c+(1-e^{-s})u      &0
    \end{pmatrix}.
\end{equation}
The tilted master operator $\tilde{\mathbb{W}}_{s}$ preserves the probability since $\sum_{x} \tilde{\mathbb{W}}_{s} (x, x') =0$ and it can describe tilted dynamics with $\exp (\tau_\mathrm{obs} \tilde{\mathbb{W}}_{s}) \left| \mathrm{S} \right>$.

Now we can calculate the marginal probability density of $K$, $\rho_{s} (K)$ and its mean value $\left< K \right>_{s}$ from Eq. (7c) in the main text:
\begin{subequations}
    \begin{align}
        \rho_{s} (K)
        &= \frac{c + (1 - e^{-s}) u}{u + c} \left( \frac{u e^{-s}}{u + c} \right)^{K},
        \\
        \left< K \right>_{s}
        &= \sum_{K} K \rho_{s} (K) = \frac{ue^{-s}}{c + (1-e^{-s})u},
    \end{align}
\end{subequations}
At $s=s_\textrm{c}$, $c + (1-e^{-s})u$ becomes zero in $\tilde{\mathbb{W}}_{s}$, and the system can be reduced to a two-state system, $\ce{S} \ce{<=>[\it{b}][\it{u + c}]} \ce{C}$, which  generates $K$ dynamical events, each one at every $\tau (= \mu^{-1} + \nu^{-1})$ on average.
The Doob's transform is not vaild in $s < s_\textrm{c}$ region where $c + (1 - e^{-s}) u$ becomes negative in $\tilde{\mathbb{W}}_{s}$.
Thus, $\left< k \right>_{s}$ has a non-zero value only if $s = s_\textrm{c}$, and is equivalent with Eq. (\ref{eqn:k_inactive}) and Eq. (\ref{eqn:k_active}).
The large deviation rate function $I(k, s)$ from $\rho_{s} (K)$ is equivalent with the Legendre-Fenchel transform of the unbiased ($s=0$) rate function, $I(k)$\cite{oono_large_1989,touchette_large_2009}:
\begin{equation}
    \begin{aligned}
        -\tau_\mathrm{obs}^{-1} \ln \rho_{s} (K)
        &\sim -k \ln p + ks
        \\
        &\sim I(k) + ks.
    \end{aligned}
\end{equation}
where $p \equiv u/(u+c)$.

\end{document}